\documentclass[10pt,letterpaper]{article}

\usepackage{graphicx}
\usepackage{epstopdf}
\usepackage{opex3}
\usepackage{color}
\usepackage{url}
\usepackage{float}
\begin{document}

\title{Nanosecond Pulse Shaping with Fiber-Based Electro-Optical Modulators and a Double-Pass Tapered Amplifier}

\author{C. E. Rogers III$^{1,2*}$ and P. L. Gould$^{1}$}

\address{$^{1}$Department of Physics, University of Connecticut, Storrs, CT 06269, USA\\$^{2}$Currently with IPITEK, 2461 Impala Drive, Carlsbad, CA 92010, USA}

\email{$^*$crogers@ipitek.com} 



\begin{abstract}
We describe a system for generating frequency-chirped and amplitude-shaped pulses on time scales from sub-nanosecond to ten nanoseconds. The system starts with cw diode-laser light at 780 nm and utilizes fiber-based electro-optical phase and intensity modulators, driven by an arbitrary waveform generator, to generate the shaped pulses. These pulses are subsequently amplified to several hundred mW with a tapered amplifier in a delayed double-pass configuration. Frequency chirps up to 5 GHz in 2 ns and pulse widths as short as 0.15 ns have been realized.
\end{abstract}



\section{Introduction}
Control of light fields is vital to many applications of lasers, such as telecommunications, optical signal processing, and quantum manipulation of atomic and molecular systems. Specific examples of the latter include coherent control of molecular dynamics \cite{Assion 1998}, population transfer in multilevel systems \cite{Collins 2012,Liu 2014}, control of ultracold collisions \cite{Wright 2007, Pechkis 2011, Wright 2015}, photoassociative formation of ultracold molecules from ultracold atoms \cite{Koch 2012, Carini 2013, Carini 2015b}, and the preparation and readout of quantum information \cite{Ladd 2010}. The two properties of light fields which are usually subject to dynamic control are amplitude (or intensity) and phase (or frequency). In some cases, polarization is also of interest.

The time scale of the control depends on the application, ranging from femtoseconds for molecular dynamics to milliseconds for ultracold atoms or ions. At high speeds, manipulations are best done in the frequency domain. A standard technique uses a pulse shaper \cite{Weiner 2000}, where a short broad-bandwidth pulse, typically $<$100 fs, is dispersed with a diffraction grating, and the amplitude and/or phase of its various frequency components are independently manipulated with a spatial light modulator. The various components are then reassembled with another diffraction grating to form the shaped output pulse. The finite frequency resolution of the pulse shaper usually limits the timescales to be faster than approximately 25 ps \cite{Monmayrant 2004}, although higher resolution has been realized with a virtually imaged phased array \cite{Shirasaki 1996}. A related technique, optical arbitrary waveform generation \cite{Cundiff 2010, Scott 2010}, separates the teeth of a frequency comb, manipulates them individually, and coherently recombines them into an output waveform. Combining separate but phase-coherent laser sources has also been investigated \cite{Wu 2015}.  
At the other extreme, low-speed manipulations are most easily performed in the time domain using time-varying electronic signals. Acousto-optical modulators can act as both optical switches and frequency shifters \cite{Thom 2013}. Their switching times are typically on the order of 10 ns or slower. Frequency chirps on the 10-100 ns time scale have been realized with diode-laser injection locking combined with either current modulation \cite{Wright 2004} or electro-optical modulation \cite{Teng 2015}.

Modulations faster than 10 ns, the typical timescale for atomic or molecular spontaneous emission, are useful for coherent manipulations. These higher speeds can be realized with electro-optical devices, which exist as either intensity modulators or phase modulators. Fiber-coupled lithium-niobate waveguide devices allow low voltages to be used, and can operate at speeds reaching up into the 10’s of GHz \cite{Kanno 2010, Wang 2015}. However, their small cross sections result in optical power limitations, especially at shorter wavelengths. Bulk electro-optical devices can handle higher power, but require high drive voltages, which makes high-speed operation challenging.

We take advantage of these high-speed waveguide-based electro-optical devices to develop a nanosecond-time-scale pulse shaping system at 780 nm. Because lithium niobate is subject to photorefractive damage at this wavelength, the time-averaged power in the modulators is limited to a few mW. Therefore, we incorporate a double-pass tapered amplifier in order to provide higher output power, up to several hundred mW, for the shaped pulses. With this system, we produce pulses with widths ranging from 0.15 ns to 10 ns and frequency chirps up to 5 GHz in 2 ns. Arbitrary intensity and frequency shapes are also demonstrated. Our technique helps fill in the gap between slow ($>$10 ns), high-resolution manipulations in the time domain and fast ($<$0.1 ns), low-resolution shaping in the frequency domain. 

The paper is organized as follows. In Sect. 2, we describe both the electronic and optical components of the pulse shaping system. In Sect. 3, we present examples of the system’s capabilities. Sect. 4 comprises concluding remarks, including possible improvements to the system as well as potential applications.

\section{Pulse shaping system}
An overall schematic of the pulse shaping system is shown in Fig. 1. The starting point is an external-cavity diode laser, generating continuous light at 780 nm. This light is coupled into a polarization-maintaining optical fiber which is connected to electro-optical phase and intensity modulators arranged in series. Both modulators are driven with a high-speed arbitrary waveform generator programmed to produce the desired frequency and amplitude patterns. The modulated light emerging from the fiber is amplified by a double pass through a tapered amplifier. Following amplification, the intensity pulses are diagnosed with a fast photodiode, while the frequency chirps are measured by combining the modulated light with a reference beam and recording the resulting heterodyne signal. Further details on the various aspects of the system are given below.
\begin{figure}[!htbp]  
	\centering
	\includegraphics[width=0.5\linewidth]{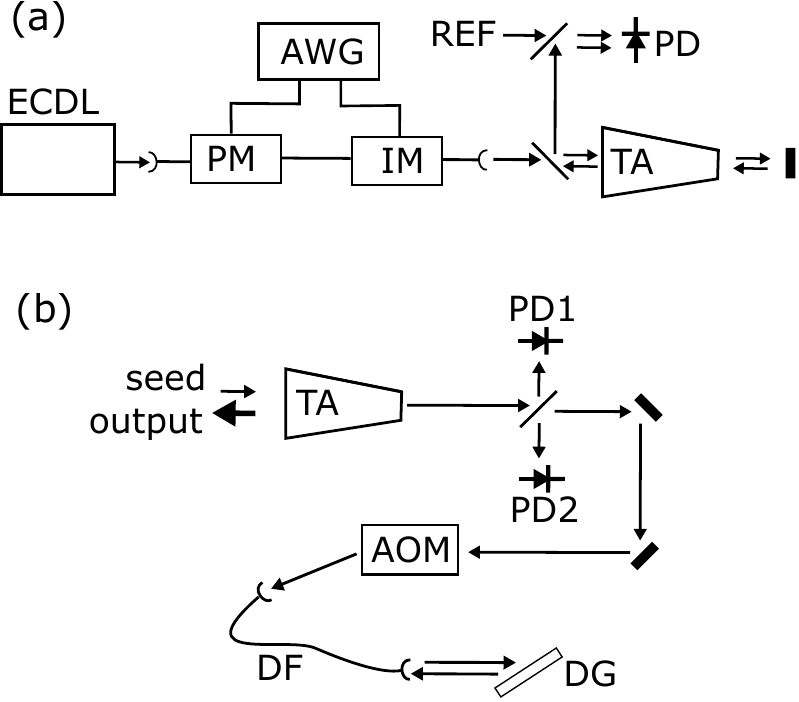}
	\caption{(a) Overall schematic of nanosecond pulse shaping system. Light from an external-cavity diode laser (ECDL) is injected, via an optical fiber, into an electro-optical phase modulator (PM), followed by an intensity modulator (IM), both of which are driven by an arbitrary waveform generator (AWG). After emerging from the fiber, the light seeds a double-pass tapered amplifier (TA). The amplified pulse is characterized by combining it with a beam from a fixed-frequency reference laser (REF) and measuring the resulting heterodyne signal on a fast photodiode (PD). (b) Details of the double-pass TA. The seed light enters the TA in the reverse direction. The light amplified in this first pass is switched by an acousto-optical modulator (AOM) into a delay fiber (DF), then reflected from a diffraction grating (DG) and returned to the TA for the second pass. A pick-off sends light to PD1 and PD2 for monitoring.}
	\label{fig:PUBFig1-v9-7-2015}
\end{figure}

The electrical signal for modulation of both the phase and amplitude is generated by a Euvis AWG801 8 GS/s one-channel arbitrary waveform generator (AWG) with 4 GHz of analog bandwidth and 11 bits of voltage resolution. In order to independently control the phase and amplitude, we employ a quasi-two-channel operation. The early part of the waveform carries the phase information, and the later part the amplitude information. The AWG has complementary outputs for its single channel, so using an rf switch and a $\sim$70 ns rf delay line, the two portions of the waveform are separately sent to their respective modulators and, after being appropriately amplified, drive them with the desired relative timing. A typical waveform contains 120 points separated by 125 ps, lasts 15 ns, and can be repeated at 10 kHz. Since the rf amplifiers are ac-coupled with low frequency cutoffs of 10-20 MHz, care must be taken to maintain the average value of the programmed waveform near zero. Also, the distortion produced by this ac coupling must be considered, especially for slowly varying waveforms. For example, a constant voltage will relax back to zero with a time constant inversely proportional to the low frequency cutoff. We compensate this distortion using an iterative algorithm.  First we measure, with an 8 GHz oscilloscope, the actual waveform driving the modulator and compare it to the desired waveform. We then adjust the programmed waveform iteratively to minimize the difference. Typically, less than 50 iterations are required for convergence.   

The phase modulator (PM) is an EOSpace model PM-0K1-00-PFA-PFA-790-S with external 50 $\Omega$ termination. Its bandwidth is specified at close to 15 GHz and the voltage (at 1 GHz) required for a $\pi$ phase shift is $V_{\pi}$ = 1.8 V. Therefore, with the 14 V maximum voltage swing available from its final amplifier (Hittite HMC-C075), we can realize a maximum phase shift of $\phi_{max}=$  7.8$\pi$. In programming the PM waveform, we must consider that the frequency shift is proportional to the time derivative of the phase: $\Delta f =\frac{1}{2\pi}\frac{d\phi}{dt}$. For example, in order to produce a linear frequency chirp, we need a quadratic variation of phase (and therefore voltage): V(t) = $\alpha t^{2}$. Assuming that the voltage varies quadratically from its maximum to its minimum and back, the product of chirp range $\Delta f$ and chirp time $\Delta t$ is limited to: $\Delta f\Delta t < 4\frac{V_{max}}{V_\pi} = 31 $. Interestingly, for a given chirp range, less voltage is required to execute the chirp in a shorter time. We note that for the fast chirps reported here, we are using a single pass through the PM. In earlier work with slower chirps \cite{Rogers 2007}, we needed multiple synchronized passes in order to build up the desired phase shift. This was realized with a fiber loop and self-injection locking of a diode laser.

The intensity modulator (IM), an EOSpace model AZ-0K5-05-PFA-PFA-790, is a device similar to the PM, but operates as a Mach-Zehnder interferometer with a split waveguide. The applied voltage causes a phase  difference $\Delta \phi (t) =  \frac{\pi}{2}\frac{V}{V_\pi}$ between its two arms, where $V_{\pi}$ = 1.6 V is the voltage change (at 1 GHz) required to go from minimum to maximum output power. The resulting interference modulates the output power according to $P(t) = P_{0}cos^{2}(\Delta\varphi(t)) $. This dependence must be accounted for when programming desired pulse shapes. The IM also has a bias input which allows dc control of the baseline phase difference between its two arms. We use a slow feedback loop to stabilize this temperature-dependent phase difference by maintaining a maximum output power when the IM is intended to be on. The rf switch which selects the IM control window from the single AWG channel is also used to apply an auxiliary bias voltage to the IM outside this control window. This serves two purposes: it compensates for the ac coupling of the rf amplifiers by maintaining a zero time-averaged voltage; and it sets the power to be high outside the control window in order to maintain seeding of the tapered amplifier.

Significant amplification of the modulated light is necessary in order to bring the power up to useful levels of more than 250 mW. The source of seed light is a relatively low power ($\sim$60 mW cw output) at 780 nm external-cavity diode laser, but more importantly, at our 780 nm wavelength, the PM and IM are subject to photorefractive damage for time-averaged input powers exceeding 5 mW. There is additional loss in the PM and IM as well as the fiber couplings and associated optics, resulting in less than 1 mW of modulated light. The typical 20 dB gain of a tapered amplifier (TA) is not sufficient, so we use the TA in a double-pass configuration \cite{Bolpasi 2010, Valenzuela 2012}, as shown in Fig. 1(b). The double-pass TA (DPTA) setup begins by sending the modulated light through the 2-watt TA (M2K TA-0785-2000-CM) in the reverse direction, i.e, the light enters the broad output facet and emerges from the narrow input (ridge) facet. Care must be taken to limit the input power so that the amplified, and spatially compressed, light emerging from the ridge facet is below the damage threshold of 50 mW. A pickoff and two photodiodes allow the powers exiting and entering the ridge facet to be monitored. After various beam-shaping optics, the single-pass-amplified light is switched by an acousto-optical modulator (AOM) and coupled into pair of delay fibers totaling 40 m ($\sim$ 200 ns), whose purpose is described below. Light emerging from the fiber is retroreflected from a 600 lines/mm diffraction grating in the Littrow configuration. The wavelength-selective grating is used instead of a nonselective mirror in order to reduce the effects of amplified spontaneous emission (ASE) which occurs over the broad gain curve of the TA. We note that the final output power of the system is limited by the losses, $\sim$12 dB, in the retro path. Since the total power at the ridge facet is limited to 50 mW, the input for the second pass through the TA is not sufficient to reach saturation. This lack of saturation may also increase the sensitivity to the etalon effects described below. 

Although the facets of the TA are antireflection coated, the gain of the TA is sufficiently high that etalon effects are problematic. As the injection current is slowly scanned, thereby changing the effective cavity length, the output power is seen to oscillate by up to 30\%. With the current fixed, the output power variation with seed wavelength is consistent with the estimated 10 GHz free spectral range of this effective cavity. These etalon effects are obviously detrimental when amplifying chirped light, but also cause problems even with unchirped pulses. Compensating the effects on a chirped pulse are discussed in Sect. 3. The impacts on unchirped pulses are minimized by using the delay fiber and AOM in the retro path. Without the delay of the retroreflected light, we see additional pulses in the time window of interest due to amplified reflections of the pulses from the TA facets. The purpose of the AOM is to interrupt the retro beam just before the pulse arrives, thereby turning off the self-lasing of the TA from ASE during the desired time window. The AOM is left on most of the time in order to ensure that the time-averaged TA output power is sufficiently high that a significant fraction of the electrical power is converted to light, thus avoiding thermal damage.

The shaped pulses are characterized using a fast photodiode (Thorlabs PDA8GS, 9.5 GHz bandwidth) and an 8 GHz oscilloscope (Agilent DSO80804A). Pulse shapes are directly recorded, while chirps are measured by combining the TA output with fixed-frequency reference light and recording the resulting heterodyne signals. The Wavelet Analysis Analytic Wavelet Transform (WAAWT) program from LabVIEW\cite{National Instruments 2015} produces a “scalogram”, a 2D (frequency and time) contour plot, from which the Wavelet Analysis Multiscale Ridge Detection (WAMRD) program generates a ridge diagram showing frequency vs. time. A third-order Savitzky-Golay filter \cite{White 2007} is typically applied to this ridge diagram in order to reduce high-frequency noise. This filter, with 20 side points, smooths a delta function to 0.5 ns full width at half maximum (FWHM).

\section{Results}
We first present sample pulse shapes for unchirped light. Fig. 2 shows Gaussian pulses with measured FWHMs ranging from 11.7 ns down to 0.15 ns. The peak power of the 150 ps pulse is $270 \pm 30$ mW. These pulse shapes are realized with programmed waveforms applied to the IM, accounting for its $sin^{2}(\Delta\varphi(t))$ response. We note that the measured FWHMs are typically $\sim$20\% larger than the programmed widths. It is possible that the IM is being overdriven, i.e., $V>V_{\pi}$ at the peak, but if this were the case, we would expect the peaks to be flattened and a small dip to appear. This is not observed. The Gaussian fits are quite good. Despite this unexplained broadening, any desired FWHM in the range shown can be realized with the appropriate adjustment to the programmed width. The fact that we can realize a FWHM as short as 150 ps is quite remarkable considering that the minimum time step of the AWG is 125 ps. The non-zero background is due primarily to ASE from the TA, which is reduced, but not eliminated, using a 10 nm FWHM interference filter on the output. The residual ASE should not cause problems in most applications because most of the light is far from the resonance(s) of interest.

\begin{figure}[!htbp]  
	\centering
	\includegraphics[width=0.5\linewidth]{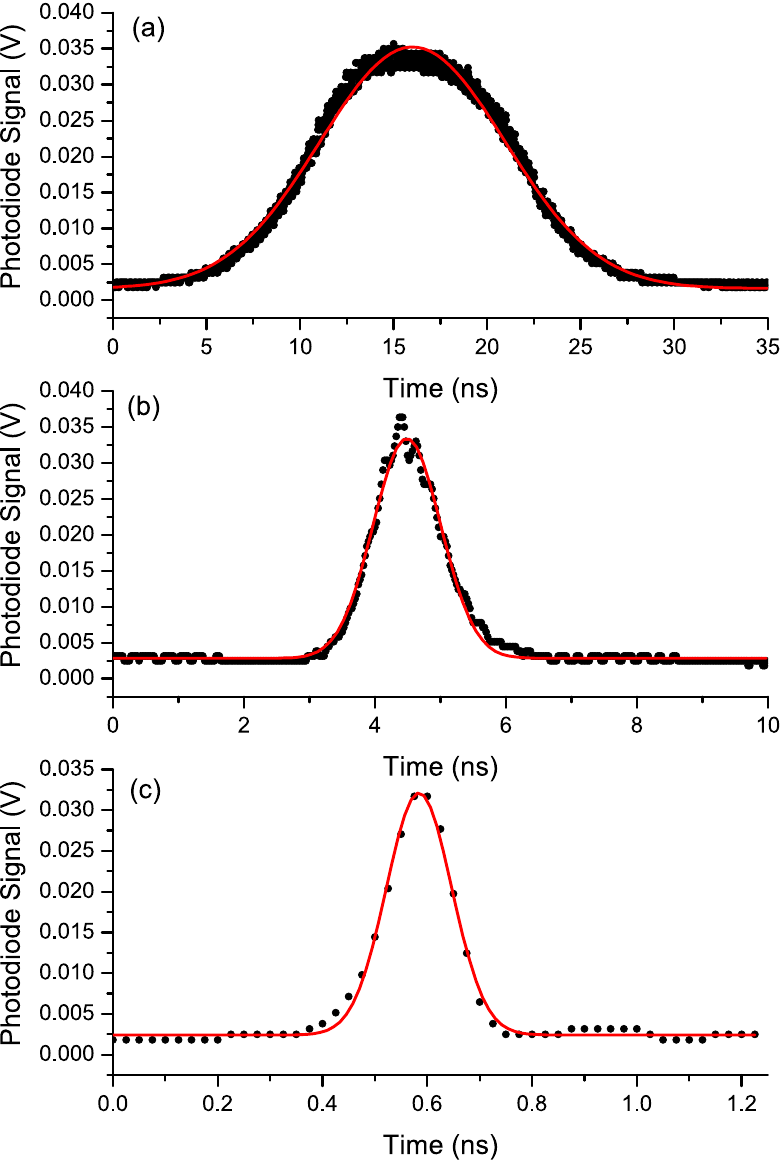}
	\caption{Gaussian pulses: measured intensity profiles (points) and Gaussian fits (solid red curves). The programmed (fitted) FWHM values are (a) 10.0 ns (11.7 ns); (b) 1.00 ns (1.22 ns); 0.10 ns (0.15 ns). The peak power of the 150 ps pulse is $270 \pm 30$ mW.}
\end{figure}
Fig. 3 shows the pulse obtained when an exponentially increasing signal, with a rapid turn-off at the peak, is applied to the IM. A fit to an exponential yields a time constant of 1.37 ns. This type of pulse shape, the time reversal of exponential decay, is relevant to experiments on efficient excitation of two-level systems with single photons \cite{Wang 2011, Aljunid 2013}.

\begin{figure}[!htbp]  
	\centering
	\includegraphics[width=0.5\linewidth]{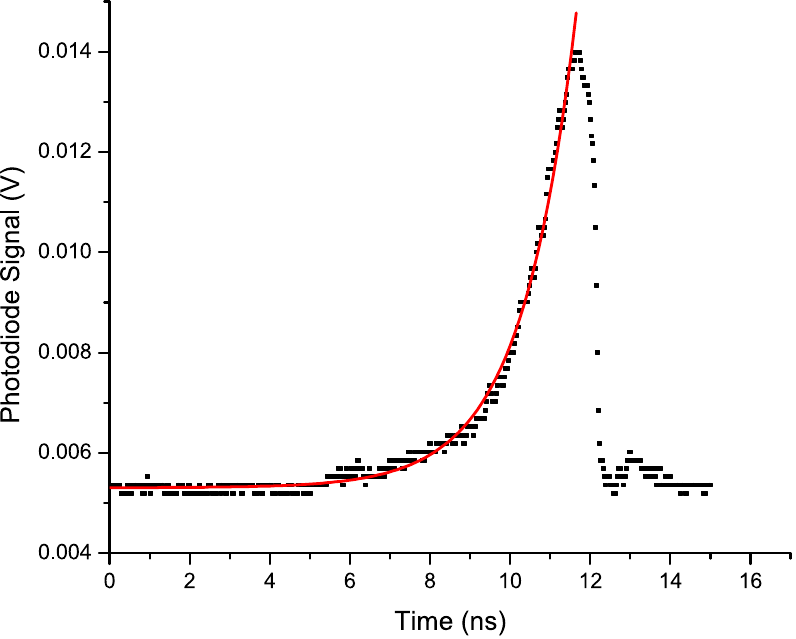}
	\caption{Exponential pulse: measured intensity (points) and exponential fit (solid red curve). The best fit time constant is 1.37 ns. }
\end{figure}

As mentioned in Sect. 2, etalon effects in the TA result in frequency chirps causing undesirable intensity modulation. However this modulation can be compensated by appropriate adjustment of the waveform applied to the IM. An example is shown in Fig. 4, where a Gaussian pulse with a linear chirp, centered about the peak of the pulse, is desired. The central frequency of the chirp coincides with a resonance of the TA cavity, resulting in symmetric dips in the intensity on either side of the peak. This coincidence is maintained by monitoring an uncompensated pulse within the same waveform, and using a servo to make small adjustments of the TA bias current. The modulated intensity pulse is compared with the desired Gaussian and, based on the difference, a correction to the AWG waveform is made. After less than 50 iterations, the shape of the compensated pulse approaches the desired one. The ultimate convergence is limited by the finite speed of the AWG.

As discussed in Sect. 2, frequency chirps are measured using a heterodyne technique. An example of a fast linear chirp, 5 GHz in 2 ns, is shown in Fig. 5. The raw heterodyne signal is depicted, as well as the resulting scalogram and the chirp (frequency vs. time) derived from the ridge of the scalogram. This chirp takes place within a 5 ns FWHM intensity pulse in order that we have adequate heterodyne signal throughout the chirp. For some applications, it may be desirable for the chirp to extend into the wings of the pulse, in which case a broader surrogate pulse must be used for the chirp diagnosis. Also shown in Fig. 5 is a smaller-range and slower linear chirp: 600 MHz in 8 ns.
\begin{figure}[!htbp]  
	\centering
	\includegraphics[width=0.50\linewidth]{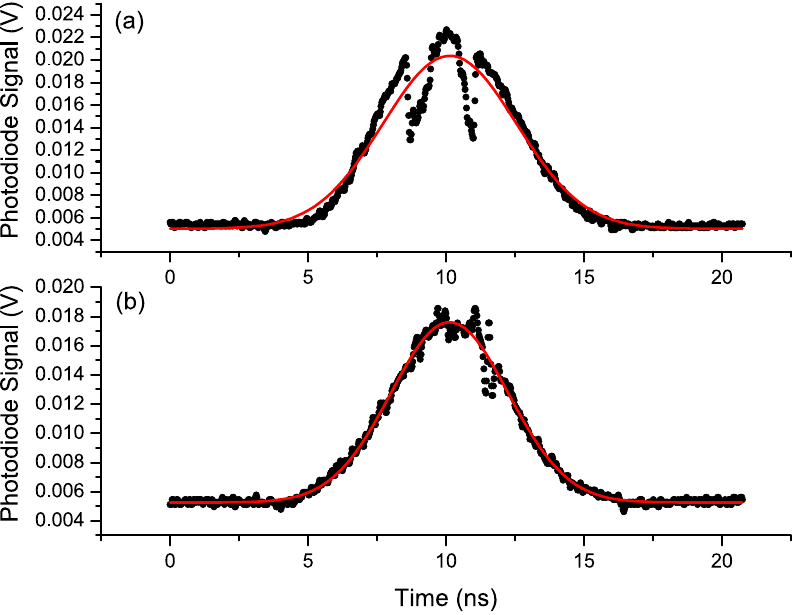}
	\caption{Compensation of cavity-mode effects. (a) Uncompensated Gaussian intensity pulse with 6 GHz in 2 ns chirp. Points are the measured intensity and the solid red curve is a Gaussian fit (FWHM = 5.7 ns). The structure in the center of the pulse is due to the frequency sweeping through a cavity resonance. (b) Compensated pulse with Gaussian fit (FWHM = 5.0 ns).}
\end{figure}
\begin{figure}[!htbp]  
	\centering
	\includegraphics[width=0.35\linewidth]{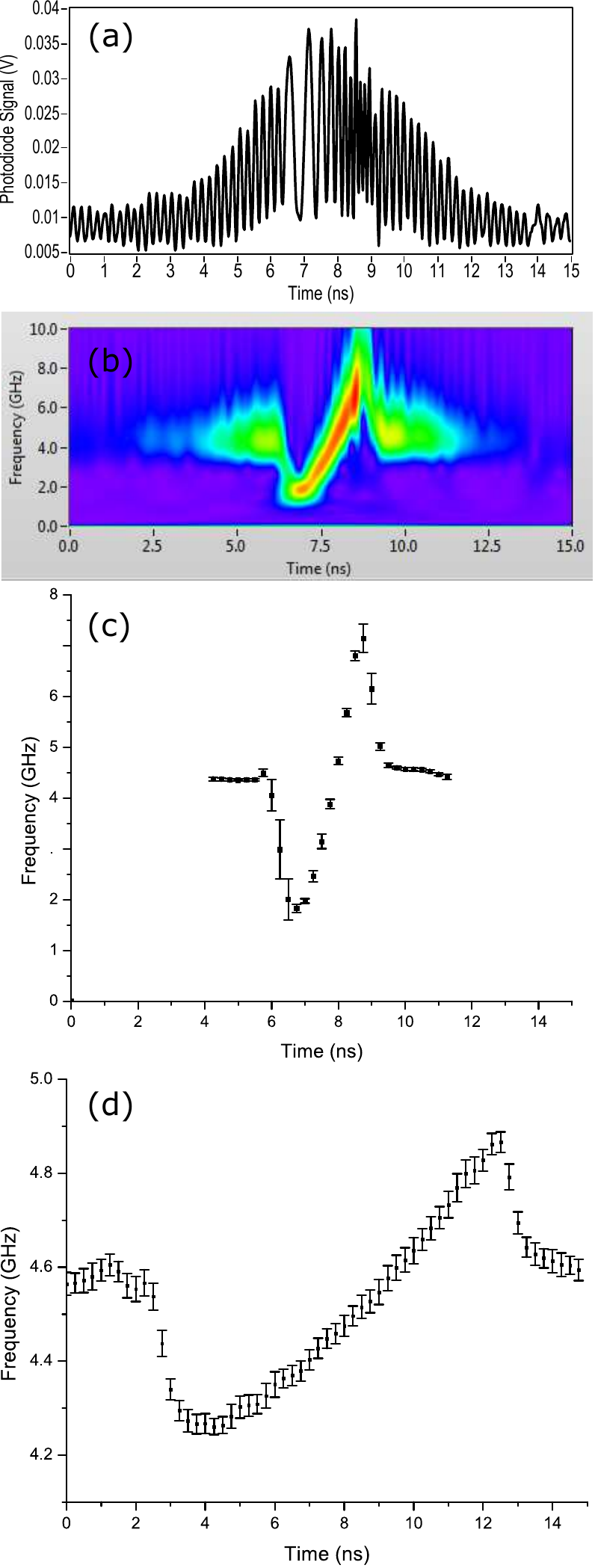}
	\caption{Measurement of frequency chirps. (a) Heterodyne signal for fast chirp. (b) Scalogram derived from (a). (c) Frequency vs. time for the fast chirp, 5 GHz in 2 ns, derived from the ridge of the scalogram in (b). (d) Frequency vs. time for slow chirp, 600 MHz in 8 ns. Error bars in (c) and (d) are statistical, based on 70 heterodyne measurements.}
\end{figure}

In Fig. 6, we display the capability of the pulse shaping system to simultaneously generate arbitrary intensity and frequency patterns. Here the target is a pair of Gaussian intensity pulses and a frequency chirp which is a sum of a linear function and an arctan function:
\begin{equation}
f(t)=(mt+b) + c\times Arctan(s(t-t_{center})).
\end{equation}
As can be seen, the measured frequency vs. time fits quite well to Eq. 1. Such a waveform would be useful for applications in adiabatic rapid passage \cite{Fetterman 1998}, Raman transitions in atoms \cite{Liu 2014}, or the photoassociative production of ultracold molecules \cite{Carini 2015}. The basic idea is to chirp slowly through the two resonances while the intensity is high, thus maintaining adiabaticity, then jump quickly between them when the intensity is low.
\begin{figure}[!htbp]  
	\centering
	\includegraphics[width=0.4\linewidth]{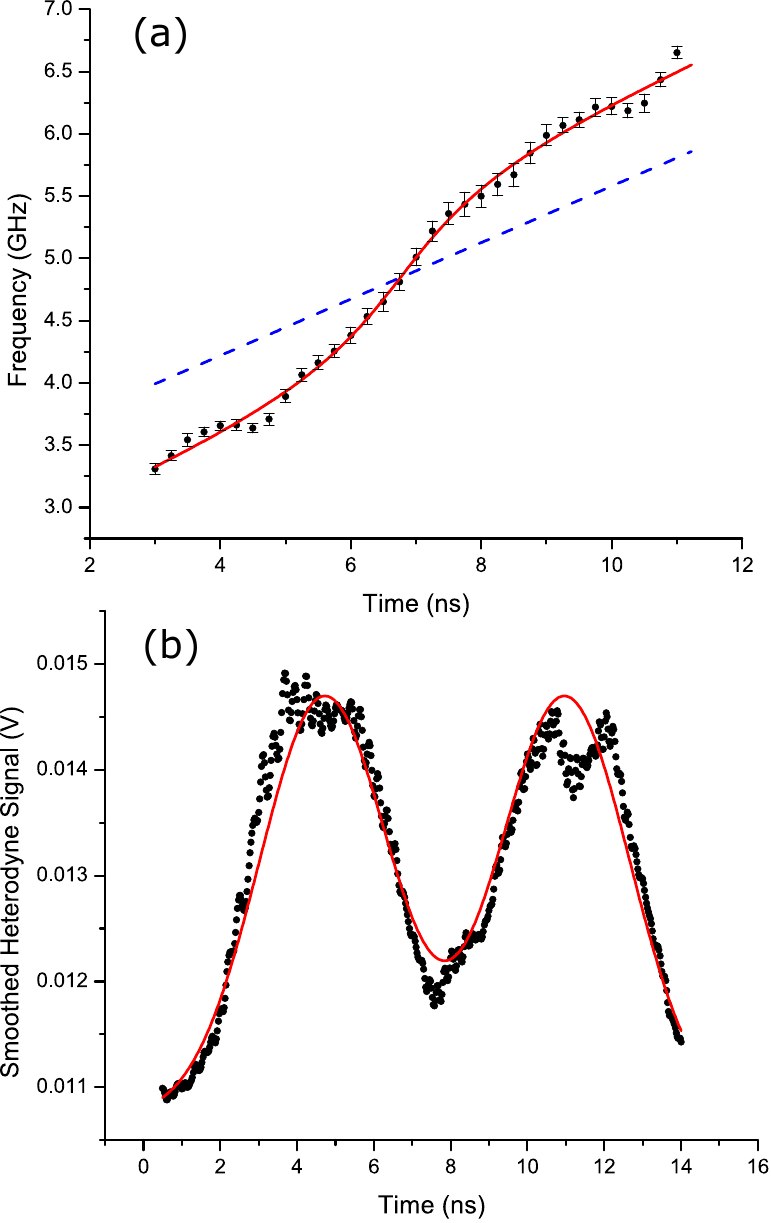}
	\caption{Arctan-plus-linear frequency chirp with double Gaussian intensity pulse. (a) Measured f(t) (points) together with fit to Eq. 1 (solid red curve). The fit parameters are: m = 0.227 GHz/ns; b = 3.31 GHz, C = 0.527 GHz; s = 0.852 ns$ ^{-1} $; $t_{center}$ = 6.77 ns. The dashed blue line is the linear contribution: f(t) = mt + b. (b) Measured I(t) (points) together with a fit to a double Gaussian (solid curve). The measurement is the result of a 1 ns smoothing of the heterodyne signal used to extract the chirp shown in (a). In the fit, the two identical Gaussians are separated by 6.25 ns, and the best fit yields FWHMs of 4.0 ns.}
\end{figure}

\section{Conclusions}
We have described the operation of a pulse shaping system for 780 nm operating on the nanosecond timescale. It is capable of producing pulses as short as 150 ps and chirps as fast as $2\times10^{18}$ Hz/s with peak powers of $>250$ mW. Although we have operated at 780 nm, the concept should work at any wavelength for which electro-optical modulators and tapered amplifiers are available. In fact, operation at longer wavelengths would have the advantage that the damage threshold for lithium niobate is higher, potentially allowing single-pass operation of the TA, as well as higher output power.  Several aspects of the system could be improved. The speed, currently limited by the electronics, could be improved with a faster AWG. Losses in the optical path, especially the retro path for the TA, could be reduced by using higher quality optics, resulting in higher output power. The output power could be dramatically increased, at the expense of a reduced repetition rate, by using a multipass Ti:sapphire amplifier in place of or in conjunction with the TA \cite{Seiler 2005}.

This system should be useful in a number of applications, especially in atomic, molecular and optical (AMO) physics. Examples include Raman transfer with a single chirped pulse \cite{Collins 2012, Liu 2014}, photoassociative formation of ultracold molecules with shaped chirped pulses \cite{Carini 2015b, Carini 2015}, efficient excitation in two-level and multi-level systems, and studies of coherent control in the presence of dissipation, e.g., due to spontaneous emission on the $10^{-8}$ s timescale.

\section{Acknowledgements}
This work was supported by the U.S. Department of Energy Office of Science, Office of Basic Energy Sciences, Chemical Sciences, Geosciences, and Biosciences Division under Award Number DE-FG02-92ER14263. We thank Jennifer Carini for construction of the TA mount and for helpful discussions, and Brad Clarke for technical assistance.

\end{document}